\documentclass[twocolumn,showpacs,prc]{revtex4}
\usepackage{graphicx}
\usepackage{dcolumn}
\usepackage{bm}
\begin{document}

\title{Biological Nuclear Transmutations as a Source of Biophotons}
\author{A. Widom}
\affiliation{Physics Department, Northeastern University, Boston MA USA}
\author{Y. N. Srivastava}
\affiliation{Physics Department \& INFN, University of Perugia, Perugia Italy}
\author{S. Sivasubramanian}
\affiliation{Nanoscale Technology and High Rate Manufacturing Research Center
\\ Northeastern University, Boston MA USA}

\begin{abstract}
Soft multi-photon radiation from hard higher energy reaction sources can be 
employed to describe three major well established properties of biophoton 
radiation; Namely, (i) the mild radiation intensity decreases for higher 
frequencies, (ii) the coherent state Poisson counting statistics, and 
(iii) the time delayed luminescence with a hyperbolic time tail. Since the soft 
photon frequencies span the visible to the ultraviolet frequency range, the 
hard reaction sources have energies extending into the nuclear transmutation 
regime. 
\end{abstract}


\maketitle

\section{Introduction \label{intro}}

{\em Biophotons} refer to a certain kind of radiation emitted by virtually 
all living systems\cite{Popp:1988,Popp:1994}. The photons are emitted in the range 
from the visible to the ultraviolet, say 
\begin{equation}
2.5\times 10^{14}\ {\rm Hz}<\left(\frac{\omega}{2\pi}\right)<10^{15}\ {\rm Hz}.
\label{intro1}
\end{equation}
The spectrum of emitted biophotons may be described by the {\em spectral distribution} 
per unit frequency per unit time per unit area, 
\begin{equation}
\frac{d^3 {\cal N}}{d\omega dt dA}\equiv \frac{d^2 \dot{\cal N}}{d\omega dA},
\label{intro2}
\end{equation}
and with a measured range of {\em total} biophoton radiation rates per unit area 
\begin{equation}
10^0\  {\rm \frac{1}{sec\ cm^2}}<\left(\frac{d \dot{\cal N}}{dA}\right)
<10^5\  {\rm \frac{1}{sec\ cm^2}}.
\label{intro3}
\end{equation}
The detailed rates depend on the biological sample being measured. 

Our purpose is to provide evidence that the biophotons are a soft photon radiation 
signature of hard higher energy nuclear transmutations in biological systems. 
Biophotons are thereby a potentially useful probe of those biological tissues which 
exhibit significant nuclear reactions.

\subsection{Biophotons Are Not Thermal \label{introA}}

The biophoton distribution is surely {\em not} what might be expected from Planck thermal 
radiation 
\begin{eqnarray}
\lambda_T=\frac{2\pi \hbar c}{k_BT}, 
\nonumber \\ 
\bar{n}(\omega )=\frac{1}{e^{\hbar \omega /k_BT}-1},
\nonumber \\ 
\lambda_T^2\left(\frac{d^2 \dot{\cal N}}{d\omega dA}\right)_T
=\left(\frac{\hbar \omega }{k_BT}\right)^2 \bar{n}(\omega ) \ \ \ {\rm (Thermal)}.
\label{intro4}
\end{eqnarray}
The exponentially infinitesimal mean number \begin{math} \bar{n} \end{math} of thermal 
photons in the experimental range Eq.(\ref{intro1}) is far too small at biological 
temperatures to produce the observed biophoton radiation distribution. The counting 
statistics of the observed biophotons are also surely {\em not} what would be 
expected from a thermal Planck-Einstein counting distribution. For  electromagnetic modes 
labeled by the index \begin{math} k \end{math}, the thermal Planck-Einstein photon 
counting statistics obey   
\begin{eqnarray}
P_T[n]=\prod_k\left[\frac{1}{1+\bar{n}_k}\left(\frac{\bar{n}_k}{1+\bar{n}_k}\right)^{n_k}\right],
\nonumber \\ 
\bar{n}_k=\bar{n}(\omega_k).
\label{intro5}
\end{eqnarray}

\subsection{Experimental Biophotons \label{introB}}

\begin{figure}[tp]
\scalebox {0.6}{\includegraphics{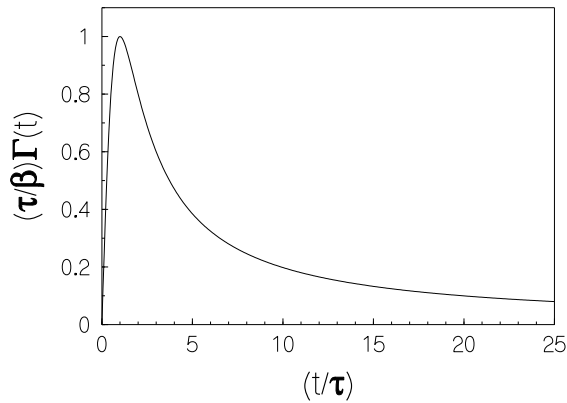}}
\caption{Shown is the theoretical prediction for the time delayed luminescence 
$\Gamma (t)$ versus time $t$ for the model Eq.(\ref{intro8}). For the soft photon 
{\em general case}, there is a hyperbolic time tail $\Gamma (t)\sim (2\beta /t)$ 
for $t \gg \tau$ wherein $\tau $ is the initial rise time.} 
\label{fig1}
\end{figure}

The experimental situation is as follows: (i) The spectral distribution of biophotons 
is a gently decreasing function of frequency\cite{Popp:1988,Popp:1994,Popp:1983}, say 
inversely with frequency 
\begin{equation}
\left(\frac{d^2 \dot{\cal N}}{d\omega dA}\right)\approx \frac{c\nu}{\omega}\ ,
\label{intro6}
\end{equation}
wherein \begin{math} \nu  \end{math} is the effective density per unit volume of biophoton 
sources. (ii) The counting 
statistics is of the Poisson type\cite{Shen:1993,Ruth:1976},
\begin{equation}
P[n]\approx \prod_k\left[\frac{\bar{n}_k ^{n_k} e^{-\bar{n}_k}}{n_k!}\right].
\label{intro7}
\end{equation}
(iii) Biophotons exhibit time delayed luminescence\cite{Strehler:1951,Li:1983,Popp:2002}. 
Such delays give rise to a time dependent transition rate with a hyperbolic long time tail; 
e.g.   
\begin{equation}
\Gamma(t)\approx \frac{2\beta t}{\tau^2+t^2}\ ,\ \ \ \ \beta={\rm coupling\ strength},
\label{intro8}
\end{equation}
plotted in what follows in FIG.\ref{fig1} and  which is inconsistent with simple 
``exponentential decay'' emission processes.
Eq.(\ref{intro8}) will be derived in Sec.\ref{TDL}.

\subsection{Soft Photon Radiation \label{introC}}

It will be shown in what follows that {\em soft} multi-photon radiation from {\em hard} 
high energy reaction sources obeying 
\begin{equation}
\Delta E \gg \hbar \omega 
\label{intro9}
\end{equation}
allow for an explanation of all three experimental properties in Eq.(\ref{intro6}), 
(\ref{intro7}) and (\ref{intro8}). Given the frequency range of biophotons in Eq.(\ref{intro1}), 
the hard high energy scale \begin{math} \Delta E \end{math} of the sources in Eq.(\ref{intro9}) 
approach those of nuclear transmutations in biological systems. This forms the basis of the theoretical 
explanation of biophotons being proposed in this work.

\section{Bio-Nuclear Transmutations \label{NT}}

We here review some extensive {\em experimental} literature on biological systems which 
describe nuclear transmutations. No detailed {\em theory} is given for how biological 
systems allow for reactions at nuclear energies beyond the statement that such large energy transformations involve collective energies of many collectively acting electrons. Many 
electrons dumping their individually smaller energies into a single high energy nuclear 
reaction may contribute to a nuclear transmutation process with plenty of available 
biophysical energy. 

Due mainly to the lack of a clear and detailed biophysical theory of nuclear transmutations, 
the experimental data is considered by some to be controversial. The present lack of a 
biophysical theory of how the collective energy of many degrees of freedom is focused on a 
few nuclear transmutation events by no means reflects badly on the reliable experimental 
data which clearly indicates nuclear transmutations. Reports of specific nuclear reactions 
which have not actually been directly observed also contribute to the controversy. If one 
does not report that which has not been directly and clearly observed, then what remains is considerable quality experimental data in favor of biophysical nuclear transmutations. 

Pioneering experimental work on biological nuclear transmutations were carried out by 
Kervran\cite{Kervran:1972}. Crabs, shellfish and crayfish have shells made largely of 
calcium. Often these animals {\em molt}, i.e. lose their shells and then create new 
shells. Typically, a crab of size \begin{math}\sim\ 17 {\rm \ cm} \times 10 {\rm \ cm} \end{math} 
is covered by a shell of weight \begin{math} \sim 350 {\rm \ gm} \end{math}. During molting, 
these animals stay away from each other since they are vulnerable. It is unlikely that 
calcium could be passed between different animals during multing. In the 
entire body of a crab there is only enough calcium to produce \begin{math} \sim 3 \% \end{math}
of a shell taking into account the calcium carbonate stored in the hepato-pancreas just 
before molting. In about a day, such a crab will grow a new shell. In water {\em completely 
devoid of calcium}, shellfish can still create their calcium-bearing shells as shown by an 
experiment performed at the Maritime Laboratory of Roscoff\cite{Kervan1}. A crayfish was 
put into a sea water basin from which calcium carbonate had been removed by precipitation. 
The animal made a new shell anyway. Chemical analysis made on animals secreting their 
shells has revealed that calcium carbonate is formed on the outer side of a membrane.  
On the opposite side of the membrane, there is no calcium\cite{Kervan1}. The detailed nuclear transmutation source of the calcium is unknown. Possibilities have been conjectured.

Normal egg shells produced by hens contain calcium. Hens were confined in an area in which 
there was no source of calcium and no calcium was present in their diet\cite{Kervan2}. 
The calcium deficiency became clearly manifested after a few days. Eggs with soft shells
were laid. Purified mica containing virtually no calcium was fed to the hens. The hens 
jumped on the mica and began scratching around it very rapidly, panting over it; then they 
rested, rolling their heads on it, threw it into the air, and began scratching it again. 
The next day eggs with normal shells (\begin{math} \approx 7{\rm gm} \end{math}) were laid. 
An experiment of this kind, using the same mica, was undertaken with guinea-fowls over a 
period of forty days\cite{Kervan2}. The administering of the mica was suspended three times 
and each time soft-shelled eggs were laid. It was suggested that the calcium in the egg 
shells was borrowed from the bones of the hens. The question of why soft eggs were laid 
when the mica was withheld remained unanswered. The detailed nuclear transmutation source 
of the calcium is again unknown.

A number of different families of microorganisms, such as Aspergillus Niger and Saccharomyces  Cerevisiae, create potassium during growth\cite{Komaki:1965,Komaki:1967}. There has also been 
a failure of balance measurements to predict actual retention of magnesium and calcium by rats. 
These were determined by direct carcass analysis\cite{Heroux:1975}. Although biological 
nuclear transmutations have certainly taken place, the detailed reactions involved are presently 
unclear. 

Bacterial cultures can have effects on nuclear reaction rates. Measurements of 
the mercury nuclear weak decay, 
\begin{equation}
^{203}_{80}Hg ~ \to ~ ^{203}_{81}Tl+e^-+\bar{\nu}, 
\end{equation} 
yield a half life of \begin{math} 46.6 \end{math} days. When such nuclei, 
embedded in \begin{math} HgCl_2  \end{math} molecules, are placed into a microbiological 
culture, the weak decay rates are severely reduced\cite{Magos:1964} indicating other 
branches of nuclear transmutation reactions. 

The bacterial culture {\em Deinococcus Radiodurans} comfortably lives, multiples and does 
not mutate under the intense radiation which does considerable damage to many inanimate 
condensed matter systems. These bacteria can live within a nuclear reactor. The detailed 
mechanism for such a possibility is unknown. However, the metals within such bacteria can 
shield radiation with sufficient surface plasma excitations associated with weak interaction 
transmutations. 

Finally, it has been experimentally shown\cite{Vysotskii:2003}, 
employing the M\"ossbauer effect, that changes in the expected isotope concentrations 
of \begin{math} ^{57}_{26}Fe  \end{math} and \begin{math} ^{55}_{25}Mn  \end{math}  
exist within growing microbial cultures. However, as an example of unfortunately reporting 
results beyond what is actually observed, we point out the final report\cite{Vysotskii:2003} 
that the strong interaction fusion was observed; i.e. 
\begin{equation}
p^+ +\  ^{55}_{25}Mn \to \  ^{55}_{26}Fe \ \ \ {\rm (strong\ fusion)}.
\label{strong}
\end{equation} 
We find it more likely that the reaction sequence takes place in the form 
\begin{eqnarray}
e^- +p^+ \to n+\nu \ \ \ {\rm (weak\ electron\ capture)}, 
\nonumber \\ 
n +\  ^{55}_{25}Mn \to \  ^{56}_{25}Mn \ \ \ {\rm (strong\ neutron\ capture)}, 
\nonumber \\ 
 ^{56}_{25}Mn \to \ ^{55}_{26}Fe +e^- +\bar{\nu } 
\ \ \ {\rm (weak\ beta\ decay)}. 
\end{eqnarray}
Altogether, 
\begin{equation}
e^- +p^+ +\  ^{55}_{25}Mn \to \  ^{55}_{26}Fe+e^- +\nu +\bar{\nu },
\label{weak}
\end{equation}
with the final state neutrinos as the signature of a weak process.
The experiment in and by itself does not distinguish between Eq.(\ref{strong}) 
and Eq.(\ref{weak}) and so we prefer workers not to report what they have not 
directly observed. The experimental data, however, remains of {\em central importance} 
as clear evidence of biological nuclear transmutations.

\section{Soft Photons \label{SPh}}

Soft photon emission from a charged particle scattering event are those photons 
radiated into coherent states which emerge from currents classical to a sufficient 
degree of accuracy\cite{Etim:1967,Kibble:1968}. As a consequence of these classical 
current sources, soft photons have Poisson counting statistics\cite{Klauder:2006} 
as in Eq.(\ref{intro7}). 

Consider a particle with charge \begin{math} e \end{math} scattering off a fixed 
target with acceleration
\begin{equation}
{\bf a}(t)=({\bf v}_f-{\bf v}_i)\delta (t)
\label{SPh1}
\end{equation}
wherein \begin{math} {\bf v}_i  \end{math} and \begin{math} {\bf v}_f  \end{math} 
are, respectively, the initial and final velocity and the delta function is 
implicitly spread out in time by the duration time \begin{math} \tau  \end{math} 
of the scattering event. The classical radiated energy is given by 
\begin{eqnarray}
{\cal E}_{rad}=\frac{2e^2}{3c^3}\int |{\bf a}(t)|^2dt,
\nonumber \\ 
{\cal E}_{rad}=\left(\frac{2e^2|{\bf v}_f-{\bf v}_i|^2}{3c^3}\right)\delta(0),
\nonumber \\ 
{\cal E}_{rad}=\left(\frac{2e^2|{\bf v}_f-{\bf v}_i|^2}{3\pi c^3}\right)
\int_0^\infty d\omega.
\nonumber \\ 
{\cal E}_{rad}=\int_0^\infty \hbar \omega dN(\omega ).
\label{SPh2}
\end{eqnarray}
If \begin{math} dN(\omega ) \end{math} is the number of photons radiated into a 
bandwidth \begin{math} d\omega \end{math}, then Eq.(\ref{SPh2}) reads 
\begin{equation}
dN(\omega )=\frac{2}{3\pi }\left(\frac{e^2}{\hbar c}\right)
\left|\frac{|{\bf v}_f-{\bf v}_i|}{c} \right|^2\frac{d\omega}{\omega }.
\label{SPh3}
\end{equation}  
For more general scattering events or decays, the distribution of emitted photons 
obeys  
\begin{equation}
dN(\omega )=\beta (\omega )\frac{d\omega }{\omega }\ \ {\rm with}
\ \ 0<\beta \equiv \lim_{\omega \to 0} \beta (\omega )<\infty .
\label{SPh4}
\end{equation}
For charged particle scattering, Eq.(\ref{SPh3}) implies 
\begin{equation}
\beta =\frac{2\alpha }{3\pi }\left|\frac{|{\bf v}_f-{\bf v}_i|}{c} \right|^2 ,
\label{SPh5}
\end{equation}
wherein \begin{math} \alpha = (e^2/\hbar c) \end{math}. The case of beta decay 
for the neutron 
\begin{equation}
n\to e^- + p^+ +\bar{\nu }\ \ {\rm yields}
\ \ \beta =\frac{2\alpha }{3\pi }\left|\frac{{\bf v}_+-{\bf v}_-}{c} \right|^2 .
\label{SPh6} 
\end{equation}
For a given nuclear reaction, the rules for calculating the quantum electrodynamic 
beta function \begin{math} \beta (\omega )  \end{math} in Eq.(\ref{SPh4}) are well 
known. Thus, the first biophoton experimental Eq.(\ref{intro6}) is explained by nuclear 
transmutations. The second biophoton experimental Eq.(\ref{intro7}) is also explained 
by nuclear transmutations since classical currents imply Poisson statistics. The third 
biophoton experimental Eq.(\ref{intro8}) can be derived but the details are a bit more 
subtle as will be shown below.

\section{Time Delayed Luminescence \label{TDL}}

In order to understand time delayed bursts of soft photons, one may consider the general 
rules of quantum mechanics concerning the notion of the decay of a quantum state. 

\subsection{Survival Amplitudes \label{SA}} 

In the general theory, let us suppose a quantum state  
\begin{math} \Psi \end{math} is present at time zero for a system with Hamiltonian 
\begin{math} H \end{math}. At time \begin{math} t \end{math} the state  
will then be \begin{math} e^{-iHt/\hbar }\Psi  \end{math}. The amplitude that the state 
\begin{math} \Psi \end{math} survives for a time \begin{math} t \end{math} is then 
\begin{equation}
{\cal S}(t)=\left(\Psi, e^{-iHt/\hbar}\Psi \right).
\label{SA1}
\end{equation}
The survival probability that the state \begin{math} \Psi \end{math} exists at time 
\begin{math} t \end{math} is the absolute value squared of the survival amplitude squared,  
\begin{equation}
P(t)=\left|{\cal S}(t)\right|^2\equiv e^{-\eta (t)}.
\label{SA2}
\end{equation}
The transition rate per unit time \begin{math} \Gamma (t) \end{math} for a transition 
away from the state 
\begin{math} \Psi \end{math} is  
\begin{eqnarray}
\Gamma (t)=\dot{\eta }(t)=2{\Re }e \dot{\chi }(t), 
\nonumber \\ 
{\rm wherein}\ \ {\cal S}(t)\equiv e^{-\chi (t)}. 
\label{SA3}
\end{eqnarray}
The special case, \begin{math} \Gamma (t)\approx \Gamma ={\rm const.} \end{math}, 
leads to an exponential decay in the survival probability 
\begin{math} P(t)\approx e^{-\Gamma t}  \end{math}. For the general case, there are 
time variations in \begin{math} \Gamma (t) \end{math} and
\begin{equation}
P(t)=\exp\left(-\int_0^t \Gamma (s)ds\right)\ .
\label{SA4}
\end{equation}

Formally, the survival amplitude of a quantum state is related to the energy 
probability density distribution of that state,
\begin{equation}
\rho(E)=\left(\Psi ,\delta (E-H)\Psi \right),
\label{SA5}
\end{equation}
via the Fourier transform 
\begin{equation}
{\cal S}(t)=\int \rho(E)e^{-iEt/\hbar }dE\equiv e^{-\chi(t)}.
\label{SA6}
\end{equation}
Eq.(\ref{SA6}) follows from Eq.(\ref{SA1}) and Eq.(\ref{SA5}).
Let us now consider the implications of these general results for the emission 
of soft photons.

\subsection{Photon Emission Rates \label{PE}}

For a state with \begin{math} n_k \end{math} photons in mode \begin{math} k \end{math}, 
the probability density in energy is given by 
\begin{eqnarray}
\rho(E)=\left<\delta \left(E-\sum_j \hbar \omega_j n_j\right)\right>,
\nonumber \\ 
\rho(E)=\sum_{[n]}
\left(\prod_k\left[\frac{\bar{n}_k ^{n_k} e^{-\bar{n}_k}}{n_k!}\right]\right)
\delta \left(E-\sum_j \hbar \omega_j n_j\right),\ 
\label{PE1}
\end{eqnarray}
wherein Eqs.(\ref{intro7}) and (\ref{SA5}) have been invoked. Eqs.(\ref{SA6}) 
and (\ref{PE1}) imply  
\begin{eqnarray}
e^{-\chi(t)}=\sum_{[n]}
\left(\prod_k\left[\frac{\bar{n}_k ^{n_k} e^{-\bar{n}_k}e^{-in_k\omega_kt}}
{n_k!}\right]\right), 
\nonumber \\ 
\chi(t)=\sum_k \bar{n}_k \left(1-e^{-i\omega_kt}\right).
\label{PE2}
\end{eqnarray}
The number of photons \begin{math} dN(\omega ) \end{math} emitted into a band 
width \begin{math} d\omega  \end{math} is given by 
\begin{equation}
dN(\omega )=\left[\sum_k \bar{n}_k \delta (\omega -\omega_k)\right]d\omega .
\label{PE3}
\end{equation}
Thus 
\begin{equation}
\chi(t)=\int_0^\infty \left(1-e^{-i\omega t}\right)dN(\omega ).
\label{PE5}
\end{equation}
The transition rate per unit time for these soft photons follows from 
Eqs.(\ref{SA3}) and (\ref{PE5}); It is 
\begin{eqnarray}
\Gamma (t)=2\int_0^\infty \omega \sin (\omega t) dN(\omega ), 
\nonumber \\ 
\Gamma (t)=2\int_0^\infty \beta (\omega ) \sin (\omega t) d\omega .
\nonumber \\ 
\beta(\omega )=\frac{1}{\pi }\int_0^\infty \Gamma (t)\sin (\omega t)dt .
\label{PE6}
\end{eqnarray}
The above Eq.(\ref{PE6}) is the central result of this section relating the 
time delayed luminescence \begin{math} \Gamma (t) \end{math} to the quantum 
electrodynamic beta function \begin{math} \beta (\omega ) \end{math}. 

As an example of such \begin{math} \beta (\omega)  \end{math} integrals in 
Eq.(\ref{PE6}), let us consider the exponential ``cut-off'' model  
\begin{equation}
\beta (\omega )=\beta e^{-\omega \tau } \ \ \ \ \Rightarrow 
\ \ \ \ \Gamma (t)=\frac{2\beta t}{\tau ^2+t^2},
\label{PE7}
\end{equation}
as in Eq.(\ref{intro8}) and plotted in FIG.\ref{fig1}. 

In the {\em general case} there is a rise time \begin{math} \tau \end{math} 
after which there is a slow hyperbolic decay in the time delayed luminescence 
\begin{math} \Gamma (t) \end{math} of the form 
\begin{equation}
\Gamma (t)\approx \frac{2\beta }{t}\ \ \ {\rm for}
\ \ \ t\gg \tau.
\label{PE8}
\end{equation}
From Eq.(\ref{PE6}) follows the rigorous limits  
\begin{equation}
\lim_{t\to \infty} t\Gamma (t)=2\lim_{\omega \to 0^+} \beta (\omega )=2\beta .
\label{PE9}
\end{equation}
Eq.(\ref{PE9}) completes our discussion of soft photon hyperbolic time delayed 
luminescence.

\section{Conclusion \label{conc}}

We have shown that soft multi-photon radiation from hard higher energy reactions 
sources can be employed to describe the three major well established properties of 
biophoton radiation. Since the soft photon frequencies span the 
visible to the ultraviolet frequency range, the hard reaction sources have 
energies extending into the nuclear transmutation regime. Thus, the biophotons 
serve as a valuable clue as to which biological systems exhibit a large number of 
nuclear transmutations.

\end{document}